\documentclass[12pt]{article}

\usepackage{graphicx}
\usepackage{rotating}

\usepackage{natbib}
\usepackage[hypertexnames=false,linktocpage=true,
breaklinks=true]{hyperref}
\usepackage{bookmark}

\hypersetup{
  colorlinks=true,
  linkcolor=blue,
  anchorcolor=blue,
  citecolor=blue,
  filecolor=blue,
  urlcolor=blue,
  bookmarksnumbered=true,
  pdfview=FitB
}

\usepackage[all]{nowidow}

\usepackage{float}

\usepackage{geometry}
\usepackage{authblk}

\newcommand\blankfootnote[1]{%
  \let\thefootnote\relax\footnotetext{#1}%
  \let\thefootnote\svthefootnote%
}

\usepackage[font=scriptsize,labelfont=bf]{caption}
\usepackage{subcaption}

\usepackage{booktabs}
\usepackage{tabularx}

\usepackage{amsmath}
\usepackage{amsfonts}
\usepackage{amssymb}

\graphicspath{{figures/}}

\begin{document}

\def\spacingset#1{\renewcommand{\baselinestretch}%
  {#1}\small\normalsize} \spacingset{1}

\title{The RITAS algorithm: a constructive yield monitor data
  processing algorithm}

\author[1]{Luis Damiano}
\author[1]{Jarad Niemi}
\affil[1]{\small Department of Statistics, Iowa State University, Ames, IA, US}
\date{December 11,2020}

\maketitle
\begin{abstract}
  Yield monitor datasets are known to contain a high percentage of
  unreliable records. The current tool set is mostly limited to
  observation cleaning procedures based on heuristic or
  empirically-motivated statistical rules for extreme value
  identification and removal. We propose a constructive algorithm for
  handling well-documented yield monitor data artifacts without
  resorting to data deletion. The four-step Rectangle creation,
  Intersection assignment and Tessellation, Apportioning, and Smoothing
  (RITAS) algorithm models sample observations as overlapping,
  unequally-shaped, irregularly-sized, time-ordered, areal spatial units
  to better replicate the nature of the destructive sampling
  process. Positional data is used to create rectangular areal spatial
  units. Time-ordered intersecting area tessellation and harvested mass
  apportioning generate regularly -shaped and \mbox{-sized} polygons
  partitioning the entire harvested area. Finally, smoothing via a
  Gaussian process is used to provide map users with spatial-trend
  visualization. The intermediate steps as well as the algorithm output
  are illustrated in maize and soybean grain yield maps for five years
  of yield monitor data collected at a research agricultural site
  located in the US Fish and Wildlife Service Neal Smith National
  Wildlife Refuge.
\end{abstract}

\blankfootnote{Creative component for the Master of Science in Statistics
\\ \href{https://dr.lib.iastate.edu/handle/20.500.12876/93760}{https://dr.lib.iastate.edu/handle/20.500.12876/93760}}

\newpage
\spacingset{2}

\section{Introduction}

As defined by the \citet{Council1997}, precision agriculture is a
data-centered discipline comprising data acquisition at an appropriate
scale, data interpretation and analysis, and management response at an
appropriate scale and time. \cite{Miller1988} were one of the firsts
to employ geostatistics explicitly in precision agriculture, a
discipline that started in the late 80's when agriculture shifted from
farm-level to site-specific crop management \citep{Oliver2010}. Their
study, among other things, introduced geostatistics for mapping
patterns in soil phosphorus or potassium via interpolation. Computer
mapping of yield and soil is one of the main uses of this technology,
typically to help customize crop management across and within fields
by identifying less productive areas at a sub-field scale
\citep{Lowenberg-DeBoer2019}. Yield data are now recorded
automatically for a wide variety of crops including cereal grains,
oilseeds, fiber, forage, biomass, fruits and vegetables. These data
are known to be accurate at a global scale, yet there exist nuances at
a local scale that affect visualization and downstream analysis, and
jeopardizes the credibility and validity of management decisions
downstream. The advent of big data has been impacting this field
largely, especially in terms of data acquisition, interpretation, and
analysis.

Numerous works have established that yield datasets contain a high
percentage of unreliable data and have proposed cleaning procedures
\citep{Blackmore1996, Moore1998, Blackmore1999, Thylen2000, Noack2003,
  Simbahan2004, Ping2005, Sudduth2007, Sudduth2012, Spekken2013,
  Leroux2018, Leroux2019, Vega2019}. Mechanical measurement errors
have been studied in great depth \citep{Arslan1999, Arslan2002,
  Grisso2002, Burks2004, Hemming2005, Fulton2009, Schuster2017}, yet
we could not find data adjustment protocols based on mechanistic
models. Typically, systematic errors are identified using either
heuristics or statistical rules with acceptable empirical results but
limited probabilistic motivation. With almost unanimous preference for
systematic error detection and removal, we have found little
development of alternative strategies for error correction or
integration such as those proposed by \cite{Bachmaier2007,
  Bachmaier2010}; in fact, some methodologies discard as much as one
third of the original dataset as summarized by \cite{Lyle2013}.

In many cases, procedures such as \citep{Vega2019} require the user to
set values for tuning parameters such as thresholds. These constants
are not learned from the data but set arbitrarily by the user,
potentially hindering comparisons across users, locations, and
years. Moreover, manually-set thresholds become a hurdle in the
context of big data: both the amount and the heterogeneity of yield
monitor datasets have been increasing as raw data become more abundant
and diverse. Additionally, the methods reviewed in \citep{Lyle2013}
have a general preference for working with crop yield as the main
input of the cleaning procedures. Being a quantity fusing data
collected from more than one sensor, each with its own propagating
measurement error, it typically has a lower signal-to-noise ratio than
the mass random variable. Also, scaling mass to homogenize
unequally-sized observational units may potentiate extreme values
resulting in a new variable with higher variability.

Some processing rules, for example the positional error removal in
\citep{Blackmore1999}, are based on the marginal distribution of yield
and fail to account for the spatial nature of the data. Others such as
\citep{Leroux2018, Vega2019} collapse all the information in one
point on a 2-dimensional plane, thus failing to recognize that the
recorded data are in fact associated with overlapping,
unequally-shaped, and irregularly-sized areal units. Each observation
is in reality a realization from a continuous spatial stochastic
process: time order and spatial superposition are intrinsic
characteristics of the destructive sampling scheme worth modeling.

The many variables logged by the yield monitor equipment include
harvested mass, geocordinates, distance traveled, and swath width,
which can be used to create an areal representation of the
data. Depending on the hardware, the reported travel distance and
speed may be measured by a speed sensor, a GPS receiver, a radar, or a
ultranosic sensor \citep{Mulla2013}. When distance is not available,
it can be linearly estimated using speed and cycle length or less
preferably approximated using the euclidean distance between two
subsequent coordinates. The swath width, also known as gathering or
cutting width, can be time dependent and the combine's path that may
not be correctly reflected in the data logged by the monitor
\citep{Ross2008}.

\cite{Blackmore1999} observed that there is a discrepancy between the
theoretical and the effective harvested area. One of the reasons is
that the recorded header width differs from the width of the header
that is truly full of crop at any given time step. While in practice it
can be set equal to a fixed proportion of the cutter bar width,
e.g. 95\%, land finishes and areas close to voids require special
treatment as the header could even be empty. In order to avoid any use
of the recorded width, and circumvent the uncertainty associated with
it, the authors introduced potential mapping. In this technique, the
recorded mass of data points belonging to circular neighbourhoods is
aggregated and assigned to a new spatial point whose location
corresponds to the circle geocenter. The aggregated mass is then
spatially re-normalized by the area of the polygons in the Voronoi
diagram formed by these new points. This approach has two
limitations. First, it transforms data into spatial points and thus
incurs in information loss about the shape and size of the
areal units (e.g. \ different shapes could have the
same centroid). Second, an additional gridding technique needs to be
applied to analyze temporal trends in yield maps
\citep{Blackmore2003}, effectively displacing the yield measurements
twice: all the polygon information is first collapsed into a single
point which is displaced afterwards.

Previously, \cite{Han1997} introduced a bitmap approach to determine
the actual combine cut width, compute the effective area size, and
approximate the new centroid of each spatial unit after removing the
area covered by previous observations. To track the covered area at
each time step, the author superimposes a regular grid where each cell
functions as a bitmask indicating whether the pixel has been harvested
before or not, hence the name of the methodology. The article does not
show how this methodology performs in the presence of sharp turns,
where overlaps tend to be larger during the pivoting motion, as these
seem to have been removed from the analysis. It is worth noting that,
even after processing, the intermediate spatial representations have
some overlap and skips left unexplained by the author. Again, as these
intermediate spatial representations are collapsed into spatial
points, some information is lost. Furthermore, these new data points
may not be spatially aligned if temporal analysis was pursued
next. Finally, the authors warn that the bitmap initialization process
is more complicated when there exists non-crop features such as
rivers, canals, or roadways.

Expanding on the bitmap concept, \cite{Drummond1999} developed a
method where observational units are represented by polygons
constructed from position and trajectory information. These are
processed in chronological harvest order by Boolean subtraction to
compute the actual harvested area during each time step, effectively
retaining full information about the areal units. As drawbacks, the
authors mention the potential overestimation at the boundaries as well
as the computational complexity of the algorithm. In the general case,
its complexity is order $O(N^2)$, yet several strategies for specific
cases are discussed there.

The principal contribution of this work is to propose a new
constructive algorithm for yield monitor data processing that accounts
for overlapping, unequally-shaped, irregularly-sized, time-ordered,
areal spatial units without resorting to data deletion. Data
collection specifics are firstly described. Then, the RITAS algorithm
is motivated and detailed on a step-by-step basis. Finally, our
methodology is illustrated with a real dataset application.

\section{Data collection}

For our purposes, we define crop yield as crop grain mass per unit
area. Our quantity of interest is a ratio of two quantities that
involve a set of typically more than six sensors, antennas, and
devices attached to combine harvesters that altogether calculate and
record grain yield in real time as the machine operator harvests the
productive field. At the end of each cycle, which lasts a
pre-specified number of seconds, the monitor logs measurements from
several on-board sensors for more than 15 variables related to
geolocation, crop characteristics, and machine diagnostics. Dataset
size could range from one to more than a hundred thousand or more
observations depending on factors such as field size, cycle length,
and land features.

Grain yield mapping using geo-referenced measurements from a
combine-mounted grain yield monitor are prevalent in
agriculture. Reported accuracy of continuous yield monitoring ranges
93-99.5\% depending on equipment type and brand, calibration regime,
flow rate, and environmental conditions at harvest
\citep{birrellComparisonSensorsTechniques1996, Fulton2009,
  Lyle2013}. \cite{Arslan2002} investigates the specific impact of
yield monitor calibration in yield estimation accuracy.

Previous agronomical studies offer insight on the physicalities of the
natural system whose sampling process we replicate. \cite{Ross2008} is
a synoptic work focused on yield estimation from yield monitor data
covering conceptual, modeling, and mechanical aspects. Grain mass
measurement involves a series of operations to separate clean grain
from other materials: gathering, cutting, pickup, and feeding,
threshing, separation, and cleaning. As a result, there is a time
delay in flow rate measurement as well as several mass losses
associated with plant gathering, processing, and leaking until the
grain is measured on or at the exit of the clean grain elevator. At
the end of each cycle, a receiver estimates the GPS antenna position,
whose placement in the combine varies across models. Their
capabilities vary largely, typically being more accurate at measuring
velocity rather than position and having a root mean square horizontal
uncertainty of 0.5-3.

\section{The RITAS algorithm}%
\label{sec:ritas}

\begin{figure}[h!]  \centering
  \includegraphics[width=\textwidth]{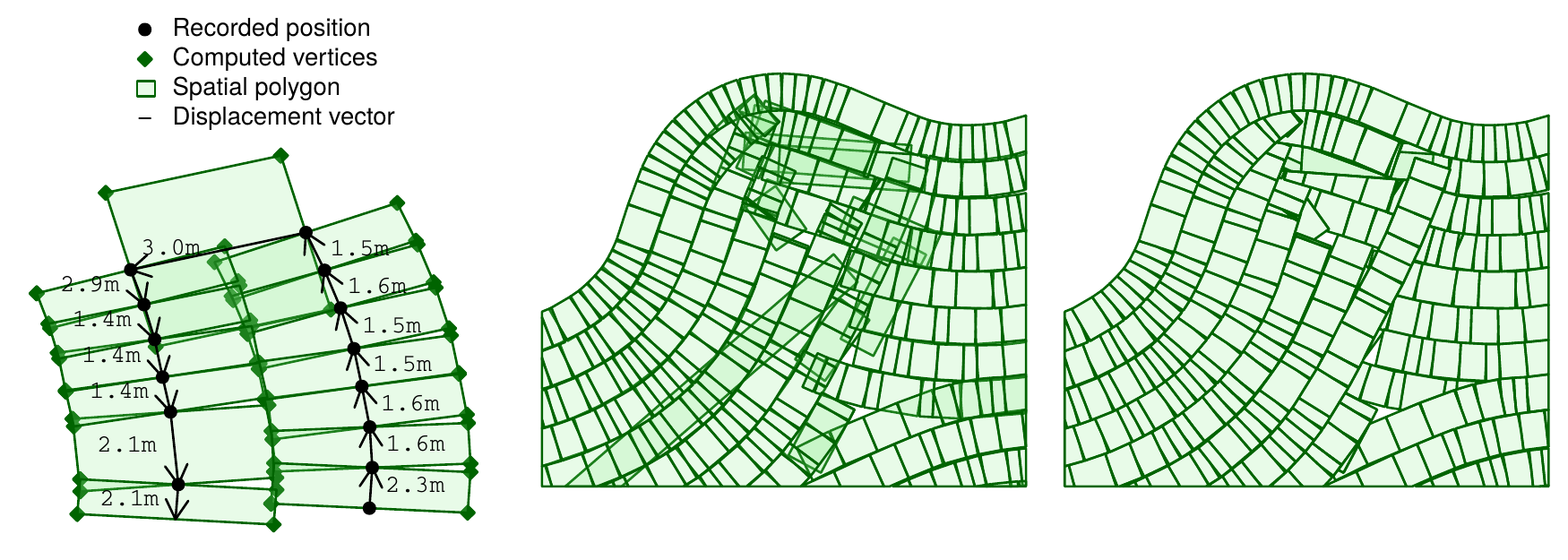}
  \caption[Close-up illustration of selected algorithm steps]{Close-up
    illustration of selected processing steps. \underline{Left}:
    Construction of the vehicle polygons from the GPS location
    data. Black dots mark the location at the end of each logging
    cycle, green dots correspond to the vertices computed according to
    the displacement vector implied by two consecutive spatial
    points. The distance traveled could be reported by the yield
    monitor or estimated from the vector length. \underline{Center}:
    Spatial polygons reveal overlap (darker areas) due to driving
    maneuvers. \underline{Right}: Tessellation eliminates the overlap
    by assigning intersecting areas to the first polygon in time.}
    \label{fig:closeup}
\end{figure}

Our algorithm constructs yield maps through the following steps:
Rectangle creation, Intersection assignment, Tessellation,
Apportioning, and Smoothing (RITAS). The overarching goal of this
process is to mimic the real world harvesting processes. Figure
\ref{fig:closeup} provides an illustration of a subset of these steps.

Each precision yield data set is assumed to have time-ordered rows
containing the following information: mass harvested $m_t$,
2-dimensional spatial coordinate $(x_t,y_t)$, and swath half-width
$w_t$ for $t=0,\ldots,T, \ T \in \mathbb{N}$.  We assume any lag time
has been effectively pre-processed by appropriately matching the mass
harvested to its 2-dimensional location. See, for example,
\cite{Burks2004, Hemming2005, Sudduth2007, Sudduth2012}.

\paragraph*{Step 1: Rectangle creation}

Figure \ref{fig:closeup} (left) illustrates the construction of a
rectangular polygon representing the harvested area between each
sequential pair of spatial coordinates. The position vector
$\mathbf{s_t} = (x_{t}, y_{t})$ represents the location of the tracked
device in a 2-dimensional plane at the time step $t$, which we assume
to be the midpoint of the combine harvester head. The rectangle is
then uniquely identified by the position of its four vertices, two
representing the beginning of the harvested area at time step $t-1$
and two representing the end of the harvested area at time step
$t$. The linear displacement vector is equal to the vector difference
between the position vectors at two subsequent time steps
$\mathbf{s_t} - \mathbf{s_{t-1}}$. The first two vertices are computed
as the endpoints of a line segment perpendicular to the displacement
vector with midpoint at $\mathbf{s_{t-1}}$ and length equal to the
swath width $2 w_t$. The remaining two vertices are found using the
same procedure but pivoting on the midpoint $\mathbf{s_t}$
instead. Given a pair of position vectors
$\mathbf{s_t}, \mathbf{s_{t-1}}$, denote
$b_t = (y_t - y_{t-1}) / (x_t - x_{t-1})$ the slope of the connecting
line, and define $dx_t = w_t (1 + b_t^{-2})^{-\frac{1}{2}}$ and
$dy_t = - dx_t b_t^{-1}$. Then, the rectangle associated with mass
$m_t$ has the four vertices coordinates given by
$\{(x_{i} \pm dx_t, y_{i} \pm dy_t): i = t, t-1\}$.

As the first spatial point has no displacement, this processing yields
$T$ rectangles with vertices collected in the set
$\mathcal{P} = \{P_{\tau}$: $\tau \in \{1, \dots, T\}\}$.

\paragraph{Step 2: Intersection assignment and Tessellation}

Figure \ref{fig:closeup} (middle) shows
end result of this construction which produces rectangles that
overlap, an area called the \emph{intersection}, due to adjacent
harvester paths. Geometries in $\mathcal{P}$ represent the area over
which the combine harvester passed, which may differ from the
effectively harvested area at each time step since the header may not
be full of crop when harvesting the field boundaries (e.g. on outer
edges, around voids), or within them. As a byproduct of the
destructive sampling scheme, the discrepancy is linked to the spatial
superposition of the observational units that arise due to local
harvesting dynamics such as turns, wedges, parallel lines, and
traveling over harvested areas. Globally within a field, this
phenomenon can be exacerbated by factors such as field
characteristics, or narrow row crops \cite{Ross2008}. Generally,
overlapping cannot be assumed symmetrical nor time-invariant. For
example, pivoting motions overlap on the inner side and the proportion
of duplicated area varies at each time step depending on factors such
as the sharpness of the turns, the angle of the wedges, or the
obstacles faced by the operator.

As a general framework to model the
time-varying effectively harvested area, we run a time-ordered
apportioning procedure over the rectangles in $\mathcal{P}$. Let
$\tilde{P}_\tau = P_t \setminus \left( \bigcup_{i = 1}^{t - 1} P_i
\right)$ be the time-ordered relative complement of the previously
harvested area in the rectangle corresponding to the time step $t$. By
doing this, we effectively map the set of overlapping rectangles
$\mathcal{P}$ into a set of non-overlapping polygons
$\tilde{\mathcal{P}} = \{\tilde{P}_{\tau}: \tau \in \{1, \dots, T
\}\}$. When the original dataset has no voids (e.g. unplanted,
flooded, or more generally inaccessible sub areas),
$\tilde{\mathcal{P}}$ form a flat plane covered by tiles with no
overlaps and no gaps (non-periodic tessellation). Some of the
computational considerations discussed in \cite{Drummond1999}, such as
the time scaling and appropriate techniques to reduce algorithm
complexity, are relevant for implementing this step.

This step creates $T$ polygons partitioning the harvested area, each
associated with the same mass it had in Step 1 but now the area may be
smaller (due to the intersection removal) and thus yield is more
accurately captured.

\paragraph*{Step 3: Apportioning} The elements in $\tilde{\mathcal{P}}$ are
unsuitable for spatial techniques that do not accommodate areas with
irregular shapes and heterogeneous area sizes. Two equally-sized long
rectangles, one in a vertical and the other in a horizontal position,
would have the same centroid yet they convey different information
about yield to the north or west of the centroids. Alternatively, two
centered equally-shaped polygons with different sizes carry different
information about the spatial coverage of the collected
data.

To normalize the areal representation in terms of both shape and size,
we superimpose a regular grid of square pixels, assign portions of the
polygons into grid pixels, and apportion the harvested mass associated
with the polygons to the pixels. The first two steps involve topology
operations on geometries whereas the last one involves manipulation of
the numerical data. Accounting for these two spatial features is by
itself a methodological improvement over the surveyed algorithms,
which simply reduce data to spatial points such as the displacement
vector endpoint or, less commonly, the polygon centroid.

Let $\mathcal{P}^{*}$ be
a set of $N \in \mathbb{N}$ equally-sized, non-overlapping, and
contiguous squared pixels forming a partition covering all the
elements in $\tilde{\mathcal{P}}$. The constant $N$ reflects the user
preference in terms of resolution, selected either by the total number
of pixels, more intuitively by the pixel length in meters, or the
target computational time investment. Although pixels size can be set
arbitrarily, its choice should consider the accuracy of the position
system. We take the pairwise intersection among the elements in
$\tilde{\mathcal{P}}$ and $\mathcal{P}^{*}$ and compute
$\pi_{\tau, n} \in [0, 1]$ the proportion of the area in the $\tau$-th
non-overlapping polygon $\tilde{P}_{\tau}$ that intersects with the
$n$-th pixel in the grid for $\tau \in \{1, \dots, T\}$ and
$n \in \{1, \dots, N\}$. The corresponding proportion of the harvested
mass $m_{\tau}$ associated with $\tilde{P}_{\tau}$ is assigned to
$P^{*}_n$. The total harvested mass associated with the $n$-th pixel
is given by $m^{*}_n = \sum_{\tau = 1}^{T} \pi_{\tau, n} \
m_{\tau}$. The resulting polygons in $\mathcal{P}^{*}$ resemble much
the Basic Areal Units (BAUs) as defined by \cite{Nguyen2012} in the
context of massive data fusion: fine-scale, nonoverlapping, areal
regions representing the smallest resolution at which data is
aggregated.

Two key aspects behind the
gridding strategy are worth mentioning. First, for the purpose of
apportioning, we assume that the harvested mass associated with the
spatial polygons is distributed uniformly within each unit. This is a
sensible assumption in this context as combine harvester log data in
short cycles and the spatial polygons represent small areas for the
scale of the underlying crop growth process.

Second, if one imposes regularity
and equal-shape conditions on the grid elements and also forces it to
cover all the elements in $\tilde{\mathcal{P}}$, the sum of the
pixels area may exceed the sum of the tessellated polygons area. The
excess, found at both the harvested region outer borders and the inner
voids boundaries, can be diagnosed visually with ease. It decreases as
the grid resolution, or equivalently as the total number of pixels
$N$, increases and so can be controlled at the cost of additional
computational time for topology operations and smoothing. 
Concretely, when a Gaussian Process is applied for smoothing as
described below, exact inference has time complexity in the order of
$O(N^3)$ and storage demands of of $O(N^2)$. In other words, as we
double the map resolution, we octuple the number of operations and
quadruple the need for storage.  Alternatively, one could exclude the
pixels with less than an arbitrary proportion of area effectively
covered by the tessellated observations. A sensible choice, such as a
minimum coverage of 50\%, will tend to balance under and over-covered
polygons. Due care should be taken so that apportioning is still
applied validly: mass should be allocated to pixels only in proportion
of the actual overlapping area, discarding any part of the tessellated
polygons that is not covered by any pixel. Note that the grid
resolution serves only for the purpose of spatial aggregation, and
need not be the same resolution used for the visualization that will
be introduced in the following step.

This step creates $N$ polygons, determined by the user based on
tesselation resolution, partitioning the harvested area each with an
associated mass of harvested crop. Figure
\ref{fig:basswood2012-main-steps} (bottom left) shows the regular grid
with apportioned mass. The regular tesselation provides constant areas
and meaningful centroids, and therefore we can use standard spatial
smoothing techniques directly on mass, as opposed to yield.

\paragraph{Step 4: Smoothing} \cite{McCullagh2006} provided empirical
evidence suggesting that the non-anthropogenic spatial variation in
yield, defined as patterns that cannot be explained by topography or
human intervention, matches the characteristics of the de Wijs process
plus white noise. We smooth using a Gaussian Process (GP) with a
Mat\'ern covariance on the logarithm of mass
\citep{handcock1993bayesian,gutt2006studies}, which becomes similar to
a de Wijs process as the length scale tends to zero. Compared with the
more common powered exponential covariance functions, e.g. Gaussian or
exponential, the Mat\'ern adds an additional parameter that controls
local smoothness, i.e. \ differentiability, and therefore is often
more accurate for real world processes.

Covariance parameters are estimated via Maximum Likelihood, and
smoothed values are found for each tile following \cite{Cressie1993}.
Specifically, for each pixel we have a predicted mean
$\hat\mu_{\ell} \in \mathbb{R}$ and variance
$\hat\sigma^2_{\ell} > 0$ for the logarithm of mass.  We using the
following formulas to convert back to the mean and variance of mass
\[ \hat{\mu}_{m} = \exp\left(\hat{\mu}_{\ell} +
\hat{\sigma}^2_{\ell}/2\right), \quad\mbox{and}\quad
\hat{\sigma}^2_{m} = \exp\left(2 \hat{\mu}_{\ell} +
\hat{\sigma}^2_{\ell}\right)
\left[\exp\left(\hat{\sigma}^2_{\ell}\right) - 1\right].
 \] Finally, yield is calculated by dividing the tile mass by the tile
area.

We have implemented our algorithm in the R programming language, using
the R packages doParallel, foreach, gstat, rgeos, and sp
\citep{Pebesma2004, Pebesma2005, Bivand2013, Graeler2016,
  Microsoft2017, Corporation2018, RCT2019, Bivand2019}.

\section{STRIPS yield mapping}

In this section, we illustrate the
functioning and the results of our methodology when applied to yield
monitor data collected from the same agricultural site over the
years. We start with a detailed discussion of the production of the
yield map for one specific year, and we finally show some resulting
visualizations for the same fields but different crops and years.

The data arises from a study
conducted at the Neal Smith National Wildlife Refuge in central Iowa
to quantify the impact of grassland-to-cropland conversion on
nitrate-nitrogen (NO\textsubscript{3}–N) concentrations in soil and
shallow groundwater and to assess the potential for perennial filter
strips to mitigate increases in NO\textsubscript{3}–N levels, run-off
reduction, and soil nutrient loss \citep{Zhou2010}. The experiment was
run in different study sites within the 3000-ha area managed by the
U.S. National Fish and Wildlife Service, located in the Walnut Creek
watershed in Jasper County, Iowa. In this case study, we focus on one
specific site named Basswood that is situated west to the Basswood
Trailhead.

Basswood, located at WGS84 15 N
0477097E 4598644N, has a total area of approximately 13 Ha. Nearly
81\% of the surface is cropland; most of the remaining proportion is
reconstructed prairie vegetation planted as part of the experimental
design. For the purpose of our yield maps, these areas are treated as
voids because no data were collected from that surface. Cropland in the
experiment is in a maize–soybean rotation using standard no-till soil
and weed-management techniques. Geographic coordinates, sample time,
moisture content, and maize (2008, 2010, 2012, 2014) and soybean (2009,
2011, 2013, 2015) flow rate were reported by a Case IH AFS Pro-600
combine-mounted yield monitor every 1-3 s during crop harvest,
resulting in a fine-scale spatially referenced dataset of crop yields
across the study area. \cite{Schulte2017} provides a detailed account
of the experiment protocol and the resulting improvements in the
biodiversity and the delivery of multiple ecosystem services.

The yield monitor dataset for the year 2012
has 4,239 observations logged every three seconds starting from the
southwest corner of the site. The swath width was reported to be
constant at 6.10m. The distance traveled during each cycle, with
an overall mean of 3.7 m, has three modes with centers at 1.9, 4.0,
and 6.0 m. The distribution of the yield reported by the monitor, with
its median located at 6.3 mg/ha and the 90\% of observations being
within 1.4 and 10.8 mg/ha, is symmetric and platykurtic. Visual
inspection suggests that there are approximately 20 extreme values on
the right tail. Small areas within the field boundaries without data,
visualized as voids in the maps, correspond to small portions of soil
allocated for nonproductive purposes (e.g. perennial crops, research
equipment).

\begin{figure}[ht]  \centering
  \includegraphics[width=\textwidth]{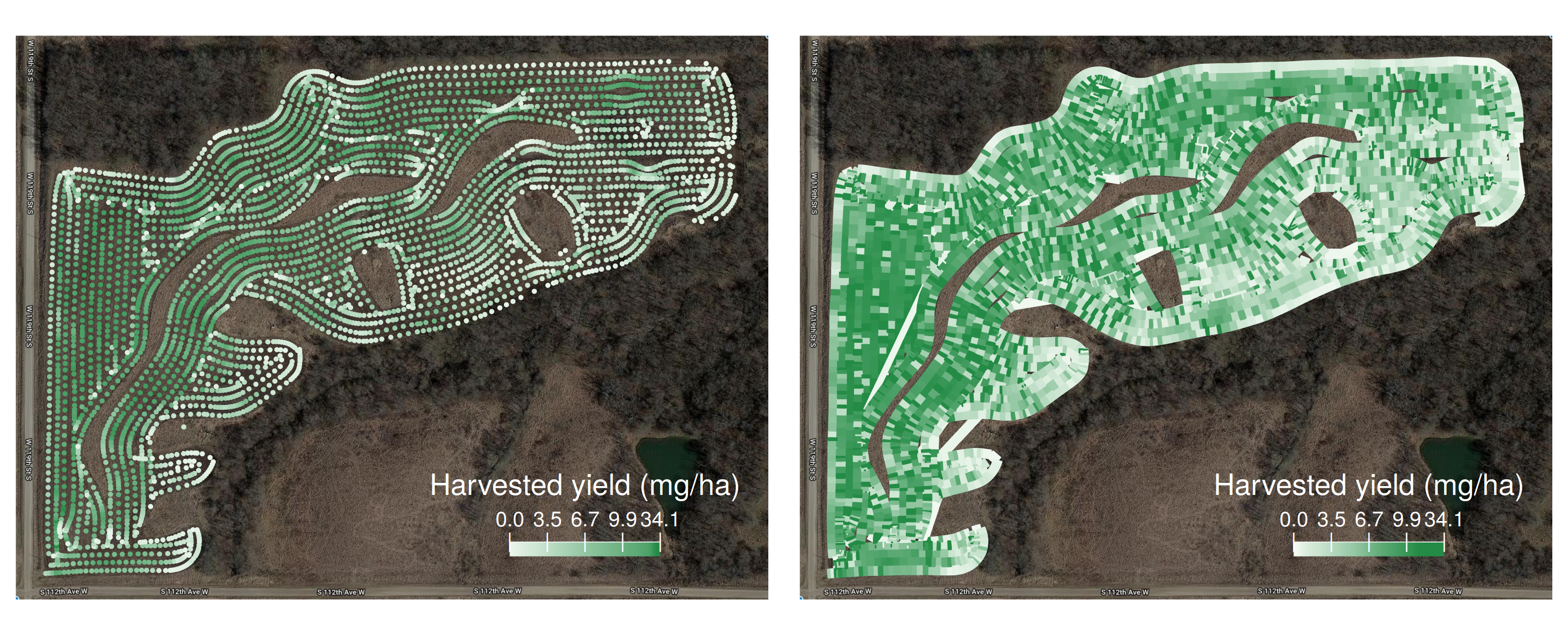}
  \includegraphics[width=\textwidth]{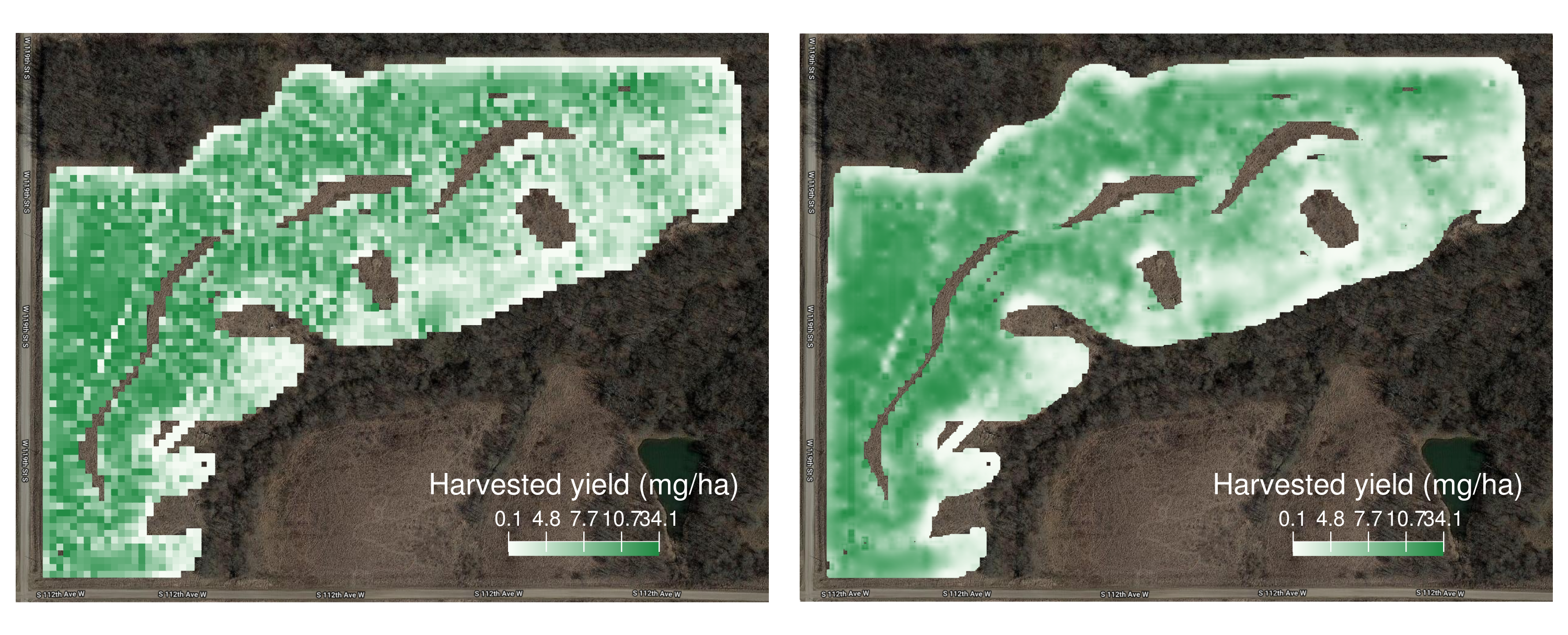}
  \caption[Visualization of selected algorithm steps as applied to a
  specific dataset]{Visualization of selected intermediate steps
    involved in the algorithm for the harvested grain yield of
    Basswood on year 2012 (maize). \underline{Top left}: Point map
    where each observation is marked with equally-sized points placed
    at the each logging location. Information relevant to spatial
    trends are hard to observe, such as area coverage and
    overlaps. \underline{Top right}: Map of the clipped
    polygons. Highly intertiment coloring in noisy areas hinder the
    visualization of the spatial trends, hence the need for
    smoothing. \underline{Bottom left}: Map of the aggregated grid at
    a 5 m resolution. This step produces equally-sized,
    regularly-shaped polygons suitable for spatial
    interpolation. \underline{Bottom right}: Map to be shown to the
    user. By increasing color homogeneity at a local level, low- and
    high-yield areas have a larger contrast and become easier to
    identify.}
  \label{fig:basswood2012-main-steps}
\end{figure}

Figure \ref{fig:basswood2012-main-steps}
(top left) displays the simplest form of a yield map. Data points are
visualized as equally-sized symbols with colors indicating the yield
value at each recorded location. A single-hue shade of green is chosen
to ease the interpretation with dark shades being intuitively
associated with higher crop yield. The characteristics of the yield
distribution vary largely across sites, crop types, and years. Since
the main objective of yield map analysis is to identify sub-field
areas with different performance levels, the points are colored
according to a measure of relative yield within the dataset as opposed
to an absolute scale. As a general rule, we define the color gradient
as a function of the the empirical quartiles in order to guarantee
that low and high yield measurements are uniformly represented at the
same time that the interpretation guidelines remain consistent across
maps.

The proposed coloring scheme helps
capturing the large-scale spatial trends: the east area, both above
and below the perennial crop strip, and the northwest area show the
best relative performance whereas the southwest area is
underperforming. Another evident pattern is outer borders and borders
around inner voids displaying lower yield, which could be explained by
soil fertility properties or could simply be a byproduct of how data
is collected. Data artifacts due to narrow finishes are well
documented: if the header cut was not full when harvesting the
boundaries and the operator failed to manually flag it, yield would be
underestimated. Careful consideration should be given to neighboring
points that are not on the borders. Light green points on the northern
borderline are adjacent to dark green points, suggesting that this
might very well be an artifact due to deficiencies in the the data
collection procedure. On the other hand, data points on the southwest
zones are consistently underperforming, thus suggesting the existence
of an actual spatial trend.

Using points to visualize the data
provides no information about the shape of the areal observational
unit and hides the overlaps in the harvested areas, which should be
considered appropriately when computing the estimated yield at a given
spatial location. In fact, from the figure it is not evident that
there is 9.0\% of areal overlap, defined as the percent excess of the
sum of the individual rectangles area over the polygons union area.

We apply the RITAS algorithm. Figure
\ref{fig:basswood2012-all-steps} (top right), displaying the
constructed spatial polygons, makes overlapping patterns more evident:
(i) subsequent samples overlap during turns, especially on the inner
side; (ii) near voids, where the landscape requires more maneuvering;
(iii) wedges formed by perpendicular passing, for example on the
western part of the field; (iv) driving from one part to another; (v)
between parallel passings and narrow segments.

Because overlapping produces a systematic
overestimation of the effective area size, thus biasing down the
estimated yield, its correct treatment can uncover underperforming
areas. In all these cases, the coloring suggests that highly
overlapping polygons are associated with lower yields. We note,
however, that yield was computed using the theoretical polygon area
which overestimates the effective harvest area. In Figure
\ref{fig:basswood2012-all-steps} (middle left), which shows the
reshaped polygons and the yield computed with the new effective area,
we notice that some of the sub field areas with low-yield polygons now
display a better performance suggesting that this visual artifact was
indeed caused by the overlapping. As a computational note, when
producing the reshaped polygons, 10 spatial polygons whose area had
been fully harvested in previous time steps were dropped from the
dataset causing an minor leakage of 0.1\% of the total harvest mass;
in case of major situations, aggregating the mass of fully nested
geometries would be appropriate.

As the ultimate goal of the visualization
work is to support the user's decision making process, the clipped
map is not adequate. Crop management decisions at a sub-field scale
are based on spatial trends whereas the measurements are highly noisy
due to a combination of at least 10 possible types of data collection
error as discussed in \cite{Lyle2013}. For example, in Figure
\ref{fig:basswood2012-all-steps} there are zones of predominating high
and low yield contaminated with scattered observations with the
opposite color, and there are also zones with a combination where it
is difficult to identify local trends.

Smoothing is thus
typically applied. As discussed in the previous section,
off-the-shelf smoothing techniques are not suitable for
unequally-sized polygons. We superimpose a grid with 4,194 squares at
a 5-meter resolution and apportion the harvested mass of the reshaped
polygons into the corresponding pixels. As we retain those pixels with
at least 50\% of their area covered by observations only, we find
local zones with under or over coverage. Overall, the whole grid
covers 2.6\% less of the sampled surface, a discrepancy that can be
diminished by increasing the grid resolution.

The results are seen in Figure
\ref{fig:basswood2012-main-steps} (bottom right). The effect of
smoothing on the signal to noise ratio is evident from the figure. The
richness of smoothing is not only on the visualization, but also on
the interpretation of the estimated parameters. We then propose that
yield maps should not only include the spatial polygons, but also
statistical information useful for the map user. The range, for
instance, is informative for map users to better understand the scale
of the spatial effect and adjust the scale of their decisions
accordingly. 

The smooth map is consistent with
the main spatial trends observed so far. Clear patterns become more
evident now; in the mixed areas, smoothing helps not only to identify
the overall local trend but also to make internal breaks/borders more
visible. Additional features can help with interpretation include
contour lines, e.g. \ each 1 mg/ha similar to \cite{Blackmore1999}, or
color schemes based on spatial clusters.


\begin{figure}[ht]  \centering
  \includegraphics[width=0.73\textwidth]{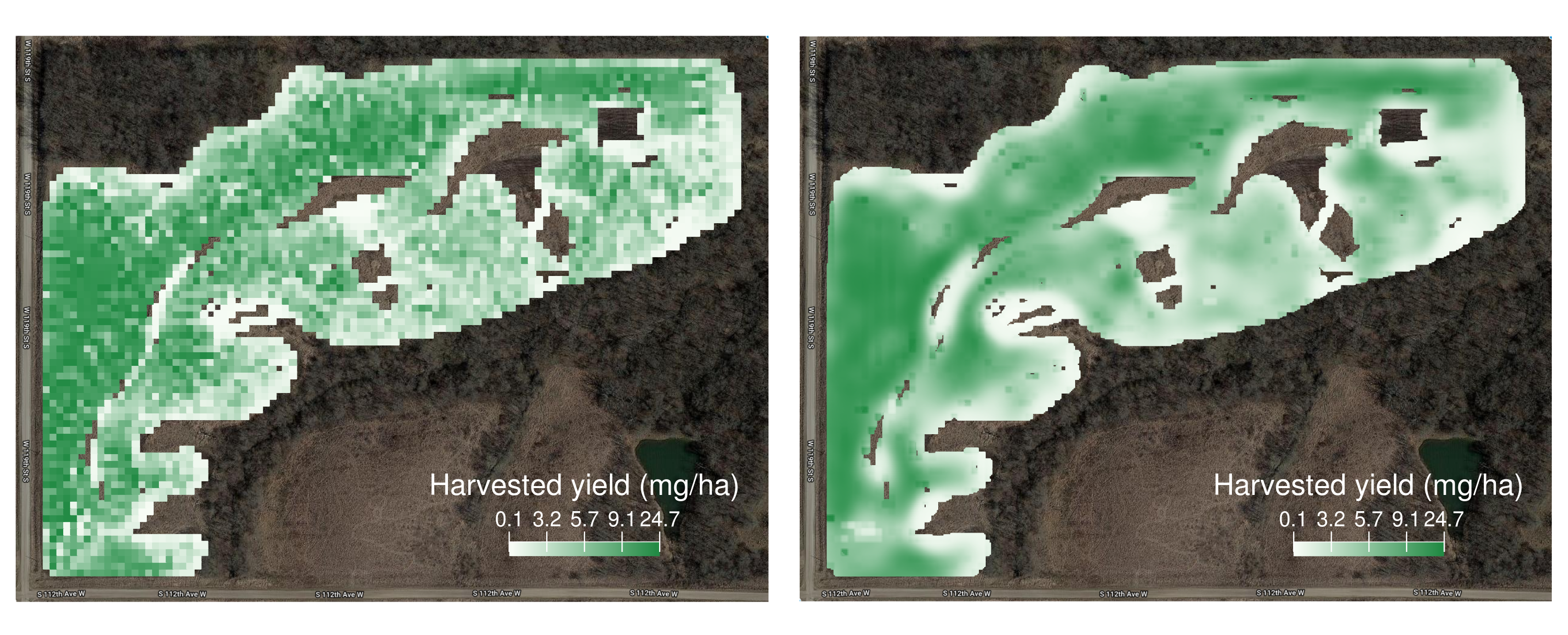}
  \includegraphics[width=0.73\textwidth]{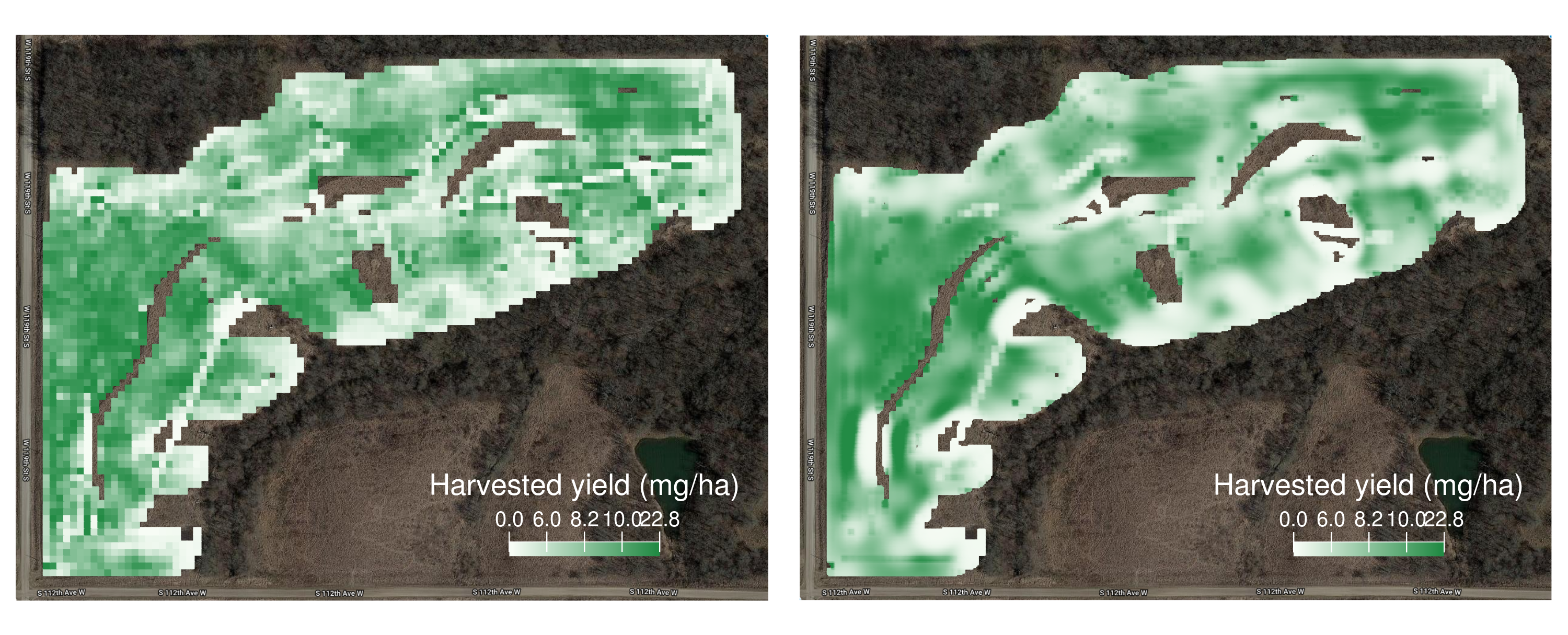}
  \includegraphics[width=0.73\textwidth]{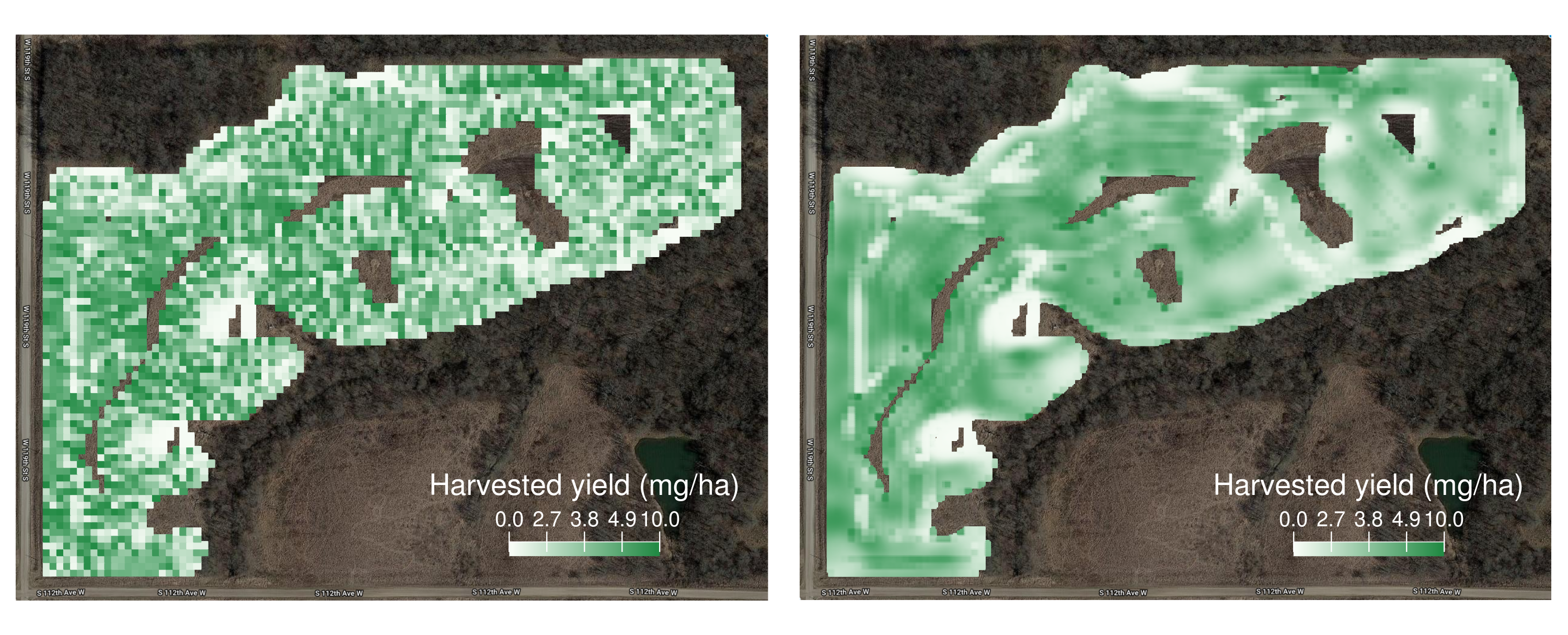}
  \includegraphics[width=0.73\textwidth]{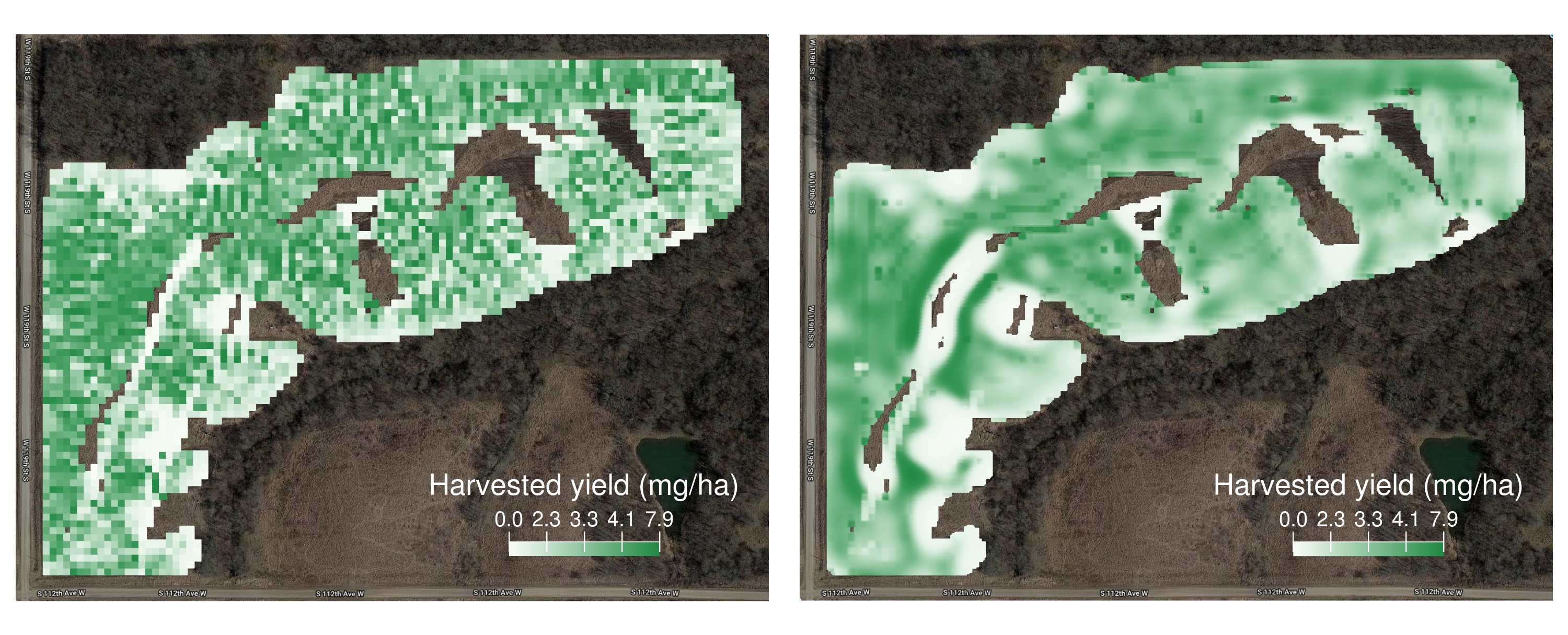}
  \caption[Visualization of the algorithm output for one field across
  four different years]{Visualization of the harvested grain yield for
    Basswood. Rows: 2010 (maize), 2014 (maize), 2009 (soybean), 2011
    (soybean). Columns: apportioned (left), smoothed (right). Despite
    soybean larger variability, the smooth map offer a clear
    visualization of the spatial trends. }%
  \label{fig:basswood-history}
\end{figure}

\section{Discussion}

We have proposed an algorithm
for data processing and visualization that is designed for situations
where the destructive sampling scheme incorporates other non-trivial
features, namely time order and unequally-sized, irregularly-shaped,
and possibly overlapping areal observational units. Reproducing the
sampling process, the rules unpack the georeferenced coordinates,
recorded as points in a 2-dimensional space, into polygons that are
next tiled. The resulting geometries are aggregated using basic areal
units and apportioned assuming that the variable of interest follows a
uniform distribution at the local scale. The final output of the
processing rules is in the form of a regular tessellation of the
harvested area, equipped with both their associated observed and
smoothed yield; the output is meant for both data analysis and
visualization. As a byproduct, because the location of the pixels in
both the aggregation and smoothing grids can be chosen freely without
affecting the validity of the algorithm, spatial registration of the
output is readily available.

To illustrate the main
characteristics of our procedure, we have presented a case study of
grain yield data processing and visualization. Whereas the previous
work in precision agriculture presents an almost unanimous preference
for extreme values detection and removal, with up to 32\% of the data
excluded \cite{Lyle2013}, our physically principled approach generates
a smooth visualization while retaining as much information as
collected. This improvement is partly explained by our reproducing of
the sampling scheme, but also by a twist on how the problem has been
traditionally framed: recognizing that local yield is a highly
volatile measurement unit due to scaling and error propagation, and
thus effectively shifting the baseline variable from yield to
mass. Also in contrast with previously proposed methodologies, most of
which require the map producer to set arbitrary thresholds, our
algorithm is autonomous. There are two main advantages associated with
this: (i) large amount of grain yield datasets can be processed
automatically, a relevant feature in the current times where data
collection has became accessible and widespread even at a small
farming scale; and (ii) the final outputs become more consistent
across different users, fields, and years.

When doing exact inference for a Gaussian Process, smoothing requires
the inversion of a square matrix with size equal to the number of
observations in the aggregation step, which in this case study is
3,738 pixels. Unless resorting to approximate interpolation methods,
significant time is expected to be consumed for this purpose. As a
palliative, our implementation optionally divides the prediction space
into smaller subsets to be computed in parallel. Although there is no
direct gain in the matrix inversion time, it provides with some
marginal improvements as out prediction space is large with more than
90,000 pixels.

Future improvements of this
algorithm could come from three perspectives: assumptions,
visualization, and computations. One of the main assumptions is that,
at a local scale, the random variable of interest follows a uniform
distribution. Although this is a reasonable idealization for the case
study given the small area of the observational units relative to the
rate of change of the underlying process, applications for
highly-distanced observations or rapidly-changing underlying processes
may require the extension to a different distribution
(e.g. exponential decay from the centroids of the polygons to its
boundaries). Yield monitors may record additional variables that are
currently not exploited by our algorithm, which could be enhanced for
example by smoothing via universal kriging with automatic relevance
determination for feature selection.

To improve visualizations,
the smooth map could profit from better display techniques to signal
the contrast between low and high valued areas (e.g. contour lines,
spatial clustering techniques). In the specific case of grain yield
maps, we still observe that some of the well-known sources of yield
data error transpire through the algorithm. Some of these can be
treated previously to running the algorithm, for example the time lag
effect, the harvester fill mode error, and position errors as
described in \cite{Blackmore1999}.

On the computational side, the
case study suggests that the most time consuming steps are smoothing
and tiling, respectively. The former could be reduced by resourcing to
approximation methods for Gaussian Process spatial interpolation
\citep{Shi2007,Cressie2008,Katzfuss2011,Nguyen2012,Nguyen2014},
approximate linear algebra routines, or a more efficient interpolation
technique. Although the intrisic sequential nature of the sampling
scheme limits the potential of parallelization in the tiling step,
some performance improvements could be potentially achieved by
subviding the spatial domain into disjoint blocks that should after be
accordingly recoupled (divide and conquer). Alternatively, the bitmap
matrix of \cite{Han1997} could be revisited as an approximation to our
polygon approach. These gains could turn the current work into a near
real-time algorithm.

\bibliography{reference/references}

\begin{thebibliography}{}

\bibitem[\protect\astroncite{Arslan and Colvin}{1999}]{Arslan1999}
Arslan, S. and Colvin, T.~S. (1999).
\newblock Laboratory performance of yield monitor.
\newblock {\em Applied Engineering in Agriculture}, 15(3):189--195.
\newblock Publisher: American Society of Agricultural and Biological Engineers
  ({ASABE}).

\bibitem[\protect\astroncite{Arslan and Colvin}{2002}]{Arslan2002}
Arslan, S. and Colvin, T.~S. (2002).
\newblock Grain yield mapping: yield sensing, yield reconstruction, and errors.
\newblock {\em Precision Agriculture}, 3(2):135--154.
\newblock Publisher: Springer Nature.

\bibitem[\protect\astroncite{Bachmaier}{2007}]{Bachmaier2007}
Bachmaier, M. (2007).
\newblock Using a robust variogram to find an adequate butterfly neighborhood
  size for one-step yield mapping using robust fitting paraboloid cones.
\newblock {\em Precision Agriculture}, 8(1):75--93.
\newblock Publisher: Springer Science and Business Media {LLC}.

\bibitem[\protect\astroncite{Bachmaier}{2010}]{Bachmaier2010}
Bachmaier, M. (2010).
\newblock A yield mapping procedure based on robust fitting paraboloid cones on
  moving elliptical neighborhoods and the determination of their size using a
  robust variogram.
\newblock {\em Positioning}, 01(1):27--41.
\newblock Publisher: Scientific Research Publishing, Inc,.

\bibitem[\protect\astroncite{Birrell
  et~al.}{1996}]{birrellComparisonSensorsTechniques1996}
Birrell, S.~J., Sudduth, K.~A., and Borgelt, S.~C. (1996).
\newblock Comparison of sensors and techniques for crop yield mapping.
\newblock {\em Computers and Electronics in Agriculture}, 14(2):215--233.

\bibitem[\protect\astroncite{Bivand and Rundel}{2019}]{Bivand2019}
Bivand, R. and Rundel, C. (2019).
\newblock rgeos: interface to geometry engine - open source ('{GEOS}').

\bibitem[\protect\astroncite{Bivand et~al.}{2013}]{Bivand2013}
Bivand, R.~S., Pebesma, E., and Gomez-Rubio, V. (2013).
\newblock {\em Applied spatial data analysis with R, second edition}.
\newblock Springer, {NY}.

\bibitem[\protect\astroncite{Blackmore and Marshall}{1996}]{Blackmore1996}
Blackmore, B.~S. and Marshall, C.~J. (1996).
\newblock {\em Yield mapping: errors and algorithms}.
\newblock American Society of Agronomy, Crop Science Society of America, Soil
  Science Society of America.

\bibitem[\protect\astroncite{Blackmore}{1999}]{Blackmore1999}
Blackmore, S. (1999).
\newblock Remedial correction of yield map data.
\newblock {\em Precision Agriculture}, 1(1):53--66.
\newblock Publisher: Springer Nature.

\bibitem[\protect\astroncite{Blackmore et~al.}{2003}]{Blackmore2003}
Blackmore, S., Godwin, R.~J., and Fountas, S. (2003).
\newblock The analysis of spatial and temporal trends in yield map data over
  six years.
\newblock {\em Biosystems Engineering}, 84(4):455--466.
\newblock Publisher: Elsevier {BV}.

\bibitem[\protect\astroncite{Burks et~al.}{2004}]{Burks2004}
Burks, T.~F., Shearer, S.~A., Fulton, J.~P., and Sobolik, C.~J. (2004).
\newblock Effects of time-varying inflow rates on combine yield monitor
  accuracy.
\newblock {\em Applied Engineering in Agriculture}, 20(3):269--275.
\newblock Publisher: American Society of Agricultural and Biological Engineers
  ({ASABE}).

\bibitem[\protect\astroncite{Corporation and Weston}{2018}]{Corporation2018}
Corporation, M. and Weston, S. (2018).
\newblock {doParallel}: foreach parallel adaptor for the 'parallel' package.

\bibitem[\protect\astroncite{Cressie and Johannesson}{2008}]{Cressie2008}
Cressie, N. and Johannesson, G. (2008).
\newblock Fixed rank kriging for very large spatial data sets.
\newblock {\em Journal of the Royal Statistical Society: Series B (Statistical
  Methodology)}, 70(1):209--226.
\newblock Publisher: Wiley.

\bibitem[\protect\astroncite{Cressie}{1993}]{Cressie1993}
Cressie, N. A.~C. (1993).
\newblock {\em Statistics for spatial data}.
\newblock John Wiley \& Sons, Inc.

\bibitem[\protect\astroncite{Drummond et~al.}{1999}]{Drummond1999}
Drummond, S.~T., Fraisse, C.~W., and Sudduth, K.~A. (1999).
\newblock Combine harvest area determination by vector processing of {GPS}
  position data.
\newblock {\em Transactions of the {ASAE}}, 42(5):1221--1228.
\newblock Publisher: American Society of Agricultural and Biological Engineers
  ({ASABE}).

\bibitem[\protect\astroncite{Fulton et~al.}{2009}]{Fulton2009}
Fulton, J.~P., Sobolik, C.~J., Shearer, S.~A., Higgins, S.~F., and Burks, T.~F.
  (2009).
\newblock Grain yield monitor flow sensor accuracy for simulated varying field
  slopes.
\newblock {\em Applied Engineering in Agriculture}, 25(1):15--21.
\newblock Publisher: American Society of Agricultural and Biological Engineers
  ({ASABE}).

\bibitem[\protect\astroncite{Grisso et~al.}{2002}]{Grisso2002}
Grisso, R.~D., Jasa, P.~J., Schroeder, M.~A., and Wilcox, J.~C. (2002).
\newblock Yield monitor accuracy: successful farming magazine case study.
\newblock {\em Applied Engineering in Agriculture}, 18(2).
\newblock Publisher: American Society of Agricultural and Biological Engineers
  ({ASABE}).

\bibitem[\protect\astroncite{Gräler et~al.}{2016}]{Graeler2016}
Gräler, B., Pebesma, E., and Heuvelink, G. (2016).
\newblock Spatio-temporal interpolation using gstat.
\newblock {\em The R Journal}, 8(1):204--218.

\bibitem[\protect\astroncite{Guttorp and Gneiting}{2006}]{gutt2006studies}
Guttorp, P. and Gneiting, T. (2006).
\newblock Studies in the history of probability and statistics {XLIX} on the
  matérn correlation family.
\newblock {\em Biometrika}, 93(4):989--995.

\bibitem[\protect\astroncite{Han et~al.}{1997}]{Han1997}
Han, S., Schneider, S.~M., Rawlins, S.~L., and Evans, R.~G. (1997).
\newblock A bitmap method for determining effective combine cut width in yield
  mapping.
\newblock {\em Transactions of the {ASAE}}, 40(2):485--490.
\newblock Publisher: American Society of Agricultural and Biological Engineers
  ({ASABE}).

\bibitem[\protect\astroncite{Handcock and Stein}{1993}]{handcock1993bayesian}
Handcock, M.~S. and Stein, M.~L. (1993).
\newblock A bayesian analysis of kriging.
\newblock {\em Technometrics}, 35(4):403--410.
\newblock Publisher: Taylor \& Francis Group.

\bibitem[\protect\astroncite{Hemming and Chaplin}{2005}]{Hemming2005}
Hemming, N. and Chaplin, J. (2005).
\newblock Determining lag time for mass flow in a combine harvester.
\newblock {\em Transactions of the {ASAE}}, 48(2):823--829.
\newblock Publisher: American Society of Agricultural and Biological Engineers
  ({ASABE}).

\bibitem[\protect\astroncite{Katzfuss and Cressie}{2011}]{Katzfuss2011}
Katzfuss, M. and Cressie, N. (2011).
\newblock Spatio-temporal smoothing and {EM} estimation for massive
  remote-sensing data sets.
\newblock {\em Journal of Time Series Analysis}, 32(4):430--446.
\newblock Publisher: Wiley.

\bibitem[\protect\astroncite{Leroux et~al.}{2018}]{Leroux2018}
Leroux, C., Jones, H., Clenet, A., Dreux, B., Becu, M., and Tisseyre, B.
  (2018).
\newblock A general method to filter out defective spatial observations from
  yield mapping datasets.
\newblock {\em Precision Agriculture}, 19(5):789--808.
\newblock Publisher: Springer Nature.

\bibitem[\protect\astroncite{Leroux et~al.}{2019}]{Leroux2019}
Leroux, C., Jones, H., Clenet, A., and Tisseyre, B. (2019).
\newblock Knowledge discovery and unsupervised detection of within-field yield
  defective observations.
\newblock {\em Computers and Electronics in Agriculture}, 156:645--659.
\newblock Publisher: Elsevier {BV}.

\bibitem[\protect\astroncite{Lowenberg-{DeBoer} and
  Erickson}{2019}]{Lowenberg-DeBoer2019}
Lowenberg-{DeBoer}, J. and Erickson, B. (2019).
\newblock Setting the record straight on precision agriculture adoption.
\newblock {\em Agronomy Journal}, 111(4):1552.
\newblock Publisher: American Society of Agronomy.

\bibitem[\protect\astroncite{Lyle et~al.}{2013}]{Lyle2013}
Lyle, G., Bryan, B.~A., and Ostendorf, B. (2013).
\newblock Post-processing methods to eliminate erroneous grain yield
  measurements: review and directions for future development.
\newblock {\em Precision Agriculture}, 15(4):377--402.
\newblock Publisher: Springer Science and Business Media {LLC}.

\bibitem[\protect\astroncite{Mark et~al.}{2013}]{Spekken2013}
Mark, S., Adriano, A., and Jose, M. (2013).
\newblock A simple method for filtering spatial data.
\newblock In {\em Precision agriculture13 (pp. 259-266)}, Precision Agriculture
  2013 - Papers Presented at the 9th European Conference on Precision
  Agriculture, {ECPA} 2013.

\bibitem[\protect\astroncite{{McCullagh} and Clifford}{2006}]{McCullagh2006}
{McCullagh}, P. and Clifford, D. (2006).
\newblock Evidence for conformal invariance of crop yields.
\newblock {\em Proceedings of the Royal Society A: Mathematical, Physical and
  Engineering Sciences}, 462(2071):2119--2143.
\newblock Publisher: The Royal Society.

\bibitem[\protect\astroncite{{Microsoft} and Weston}{2017}]{Microsoft2017}
{Microsoft} and Weston, S. (2017).
\newblock foreach: provides foreach looping construct for r.

\bibitem[\protect\astroncite{Miller et~al.}{1988}]{Miller1988}
Miller, M.~P., Singer, M.~J., and Nielsen, D.~R. (1988).
\newblock Spatial variability of wheat yield and soil properties on complex
  hills.
\newblock {\em Soil Science Society of America Journal}, 52(4):1133.
\newblock Publisher: Soil Science Society of America.

\bibitem[\protect\astroncite{Moore}{1998}]{Moore1998}
Moore, M. (1998).
\newblock An investigation into the accuracy of yield maps and their subsequent
  use in crop management.

\bibitem[\protect\astroncite{Mulla}{2013}]{Mulla2013}
Mulla, D.~J. (2013).
\newblock Twenty five years of remote sensing in precision agriculture: key
  advances and remaining knowledge gaps.
\newblock {\em Biosystems Engineering}, 114(4):358--371.
\newblock Publisher: Elsevier {BV}.

\bibitem[\protect\astroncite{{National Research Council}}{1997}]{Council1997}
{National Research Council} (1997).
\newblock {\em Precision agriculture in the 21st century}.
\newblock National Academies Press.

\bibitem[\protect\astroncite{Nguyen et~al.}{2012}]{Nguyen2012}
Nguyen, H., Cressie, N., and Braverman, A. (2012).
\newblock Spatial statistical data fusion for remote sensing applications.
\newblock {\em Journal of the American Statistical Association},
  107(499):1004--1018.
\newblock Publisher: Informa {UK} Limited.

\bibitem[\protect\astroncite{Nguyen et~al.}{2014}]{Nguyen2014}
Nguyen, H., Katzfuss, M., Cressie, N., and Braverman, A. (2014).
\newblock Spatio-temporal data fusion for very large remote sensing datasets.
\newblock {\em Technometrics}, 56(2):174--185.
\newblock Publisher: Informa {UK} Limited.

\bibitem[\protect\astroncite{Oliver}{2010}]{Oliver2010}
Oliver, M., editor (2010).
\newblock {\em Geostatistical applications for precision agriculture}.
\newblock Springer Netherlands.

\bibitem[\protect\astroncite{P et~al.}{2003}]{Noack2003}
P, N., T, M., and M, D. (2003).
\newblock {\em An algorithm for automatic detection and elimination of
  defective yield data.}
\newblock Wageningen Academic Publishers.

\bibitem[\protect\astroncite{Pebesma}{2004}]{Pebesma2004}
Pebesma, E.~J. (2004).
\newblock Multivariable geostatistics in s: the gstat package.
\newblock {\em Computers \& Geosciences}, 30:683--691.

\bibitem[\protect\astroncite{Pebesma and Bivand}{2005}]{Pebesma2005}
Pebesma, E.~J. and Bivand, R.~S. (2005).
\newblock Classes and methods for spatial data in r.
\newblock {\em R News}, 5(2):9--13.

\bibitem[\protect\astroncite{Ping and Dobermann}{2005}]{Ping2005}
Ping, J.~L. and Dobermann, A. (2005).
\newblock Processing of yield map data.
\newblock {\em Precision Agriculture}, 6(2):193--212.
\newblock Publisher: Springer Science and Business Media {LLC}.

\bibitem[\protect\astroncite{{R Core Team}}{2019}]{RCT2019}
{R Core Team} (2019).
\newblock R: a language and environment for statistical computing.

\bibitem[\protect\astroncite{Ross et~al.}{2008}]{Ross2008}
Ross, K.~W., Morris, D.~K., and Johannsen, C.~J. (2008).
\newblock A review of intra-field yield estimation from yield monitor data.
\newblock {\em Applied Engineering in Agriculture}, 24(3):309--317.
\newblock Publisher: American Society of Agricultural and Biological Engineers
  ({ASABE}).

\bibitem[\protect\astroncite{Schulte et~al.}{2017}]{Schulte2017}
Schulte, L.~A., Niemi, J., Helmers, M.~J., Liebman, M., Arbuckle, J.~G., James,
  D.~E., Kolka, R.~K., O'Neal, M.~E., Tomer, M.~D., Tyndall, J.~C., Asbjornsen,
  H., Drobney, P., Neal, J., Ryswyk, G.~V., and Witte, C. (2017).
\newblock Prairie strips improve biodiversity and the delivery of multiple
  ecosystem services from corn–soybean croplands.
\newblock {\em Proceedings of the National Academy of Sciences},
  114(42):11247--11252.
\newblock Publisher: Proceedings of the National Academy of Sciences.

\bibitem[\protect\astroncite{Schuster et~al.}{2017}]{Schuster2017}
Schuster, J.~N., Darr, M.~J., and {McNaull}, R.~P. (2017).
\newblock Performance benchmark of yield monitors for mechanical and
  environmental influences.
\newblock In {\em 2017 {ASABE} annual international meeting}, page~1.
\newblock American Society of Agricultural and Biological Engineers.

\bibitem[\protect\astroncite{Shi and Cressie}{2007}]{Shi2007}
Shi, T. and Cressie, N. (2007).
\newblock Global statistical analysis of {MISR} aerosol data: a massive data
  product from {NASAs} terra satellite.
\newblock {\em Environmetrics}, 18(7):665--680.
\newblock Publisher: Wiley.

\bibitem[\protect\astroncite{Simbahan et~al.}{2004}]{Simbahan2004}
Simbahan, G.~C., Dobermann, A., and Ping, J.~L. (2004).
\newblock Screening yield monitor data improves grain yield maps.
\newblock {\em Agronomy Journal}, 96(4):1091.
\newblock Publisher: American Society of Agronomy.

\bibitem[\protect\astroncite{Sudduth and Drummond}{2007}]{Sudduth2007}
Sudduth, K.~A. and Drummond, S.~T. (2007).
\newblock Yield editor.
\newblock {\em Agronomy Journal}, 99(6):1471.
\newblock Publisher: American Society of Agronomy.

\bibitem[\protect\astroncite{Sudduth et~al.}{2012}]{Sudduth2012}
Sudduth, K.~A., Drummond, S.~T., and Myers, D.~B. (2012).
\newblock Yield editor 2.0: software for automated removal of yield map errors.
\newblock In {\em 2012 dallas, texas, july 29 - august 1, 2012}. American
  Society of Agricultural and Biological Engineers.

\bibitem[\protect\astroncite{Thylén et~al.}{2000}]{Thylen2000}
Thylén, L., Algerbo, P.~A., and Giebel, A. (2000).
\newblock An expert filter removing erroneous yield data.
\newblock In Robert, P.~C., Rust, R.~H., and Larson, W.~E., editors, {\em
  Proceedings of the 5{\textless}span
  class="nocase"{\textgreater}$^{\textrm{th}}${\textless}/span{\textgreater}
  international conference on precision agriculture, bloomington, minnesota,
  {USA}, 16-19 july, 2000.}, pages 1--9. American Society of Agronomy.

\bibitem[\protect\astroncite{Vega et~al.}{2019}]{Vega2019}
Vega, A., Córdoba, M., Castro-Franco, M., and Balzarini, M. (2019).
\newblock Protocol for automating error removal from yield maps.
\newblock {\em Precision Agriculture}.
\newblock Publisher: Springer Nature.

\bibitem[\protect\astroncite{Zhou et~al.}{2010}]{Zhou2010}
Zhou, X., Helmers, M.~J., Asbjornsen, H., Kolka, R., and Tomer, M.~D. (2010).
\newblock Perennial filter strips reduce nitrate levels in soil and shallow
  groundwater after grassland-to-cropland conversion.
\newblock {\em Journal of Environment Quality}, 39(6):2006.
\newblock Publisher: American Society of Agronomy.

\end{thebibliography}

\section{Acknowledgements}

Funding was provided by Iowa State University through the Presidential
Interdisciplinary Research Initiative on C-CHANGE: Science for a
Changing Agriculture and the Foundation for Food and Agriculture
Research (FFAR). The dataset used to illustrate our work was collected
as part of the Science-Based Trials of Rowcrops Integrated with
Prairie Strips (STRIPS) project, which involved the work of the US Fish
and Wildlife Service Neal Smith National Wildlife Refuge, numerous
technicians and student researchers, and Heartland Cooperative.

\appendix

\begin{figure}
  \begin{subfigure}[b]{0.49\textwidth}
    \centering
    \includegraphics[width=1\textwidth]{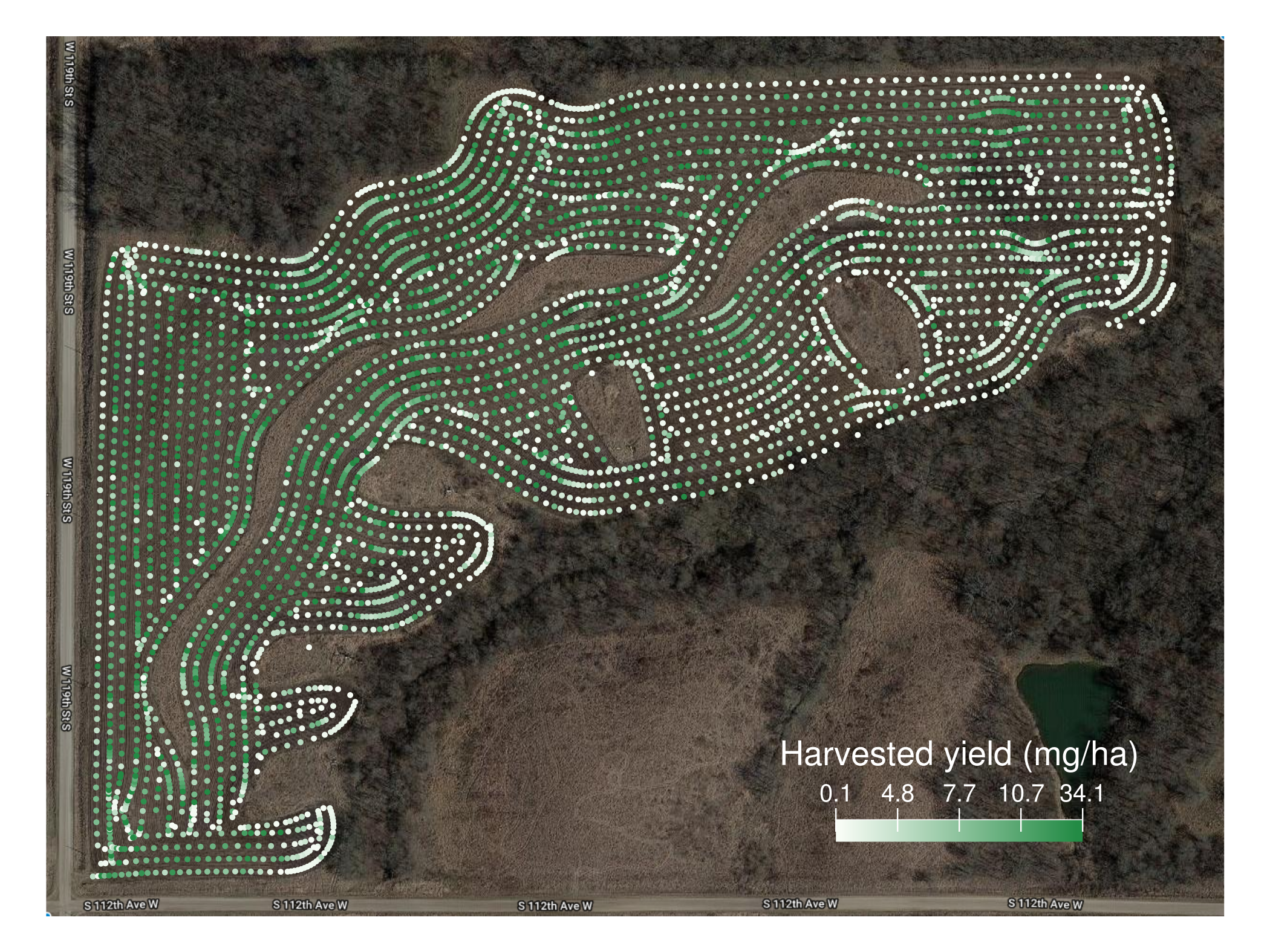}
    \caption{Observations as point on a 2D-space}
   \end{subfigure}
  \begin{subfigure}[b]{0.49\textwidth}
    \centering
    \includegraphics[width=1\textwidth]{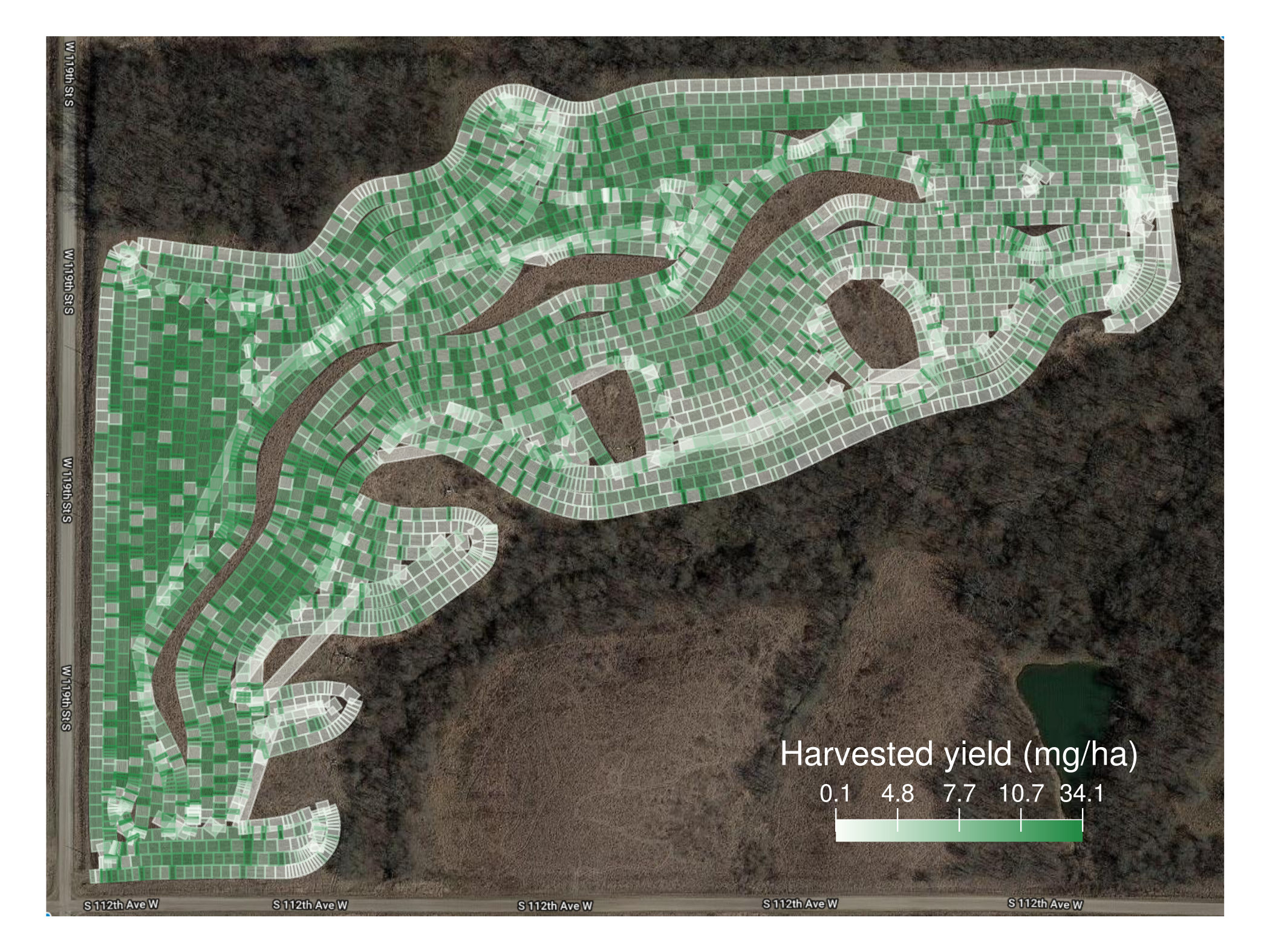}
    \caption{Rectangle creation step output}
  \end{subfigure}
  \begin{subfigure}[b]{0.49\textwidth}
    \centering
    \includegraphics[width=1\textwidth]{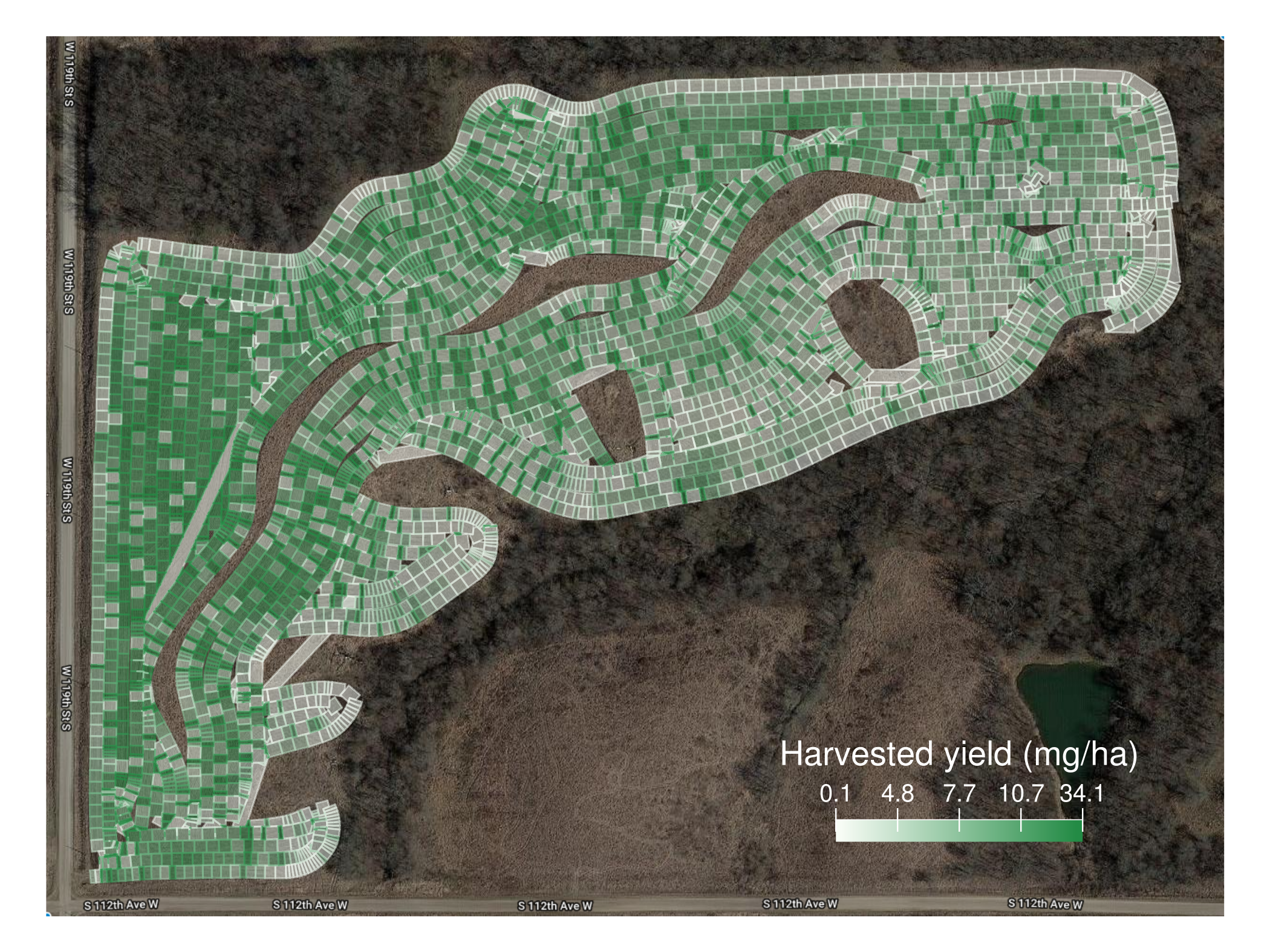}
    \caption{Tessellation step output}
  \end{subfigure}
  \begin{subfigure}[b]{0.49\textwidth}
    \centering
    \includegraphics[width=1\textwidth]{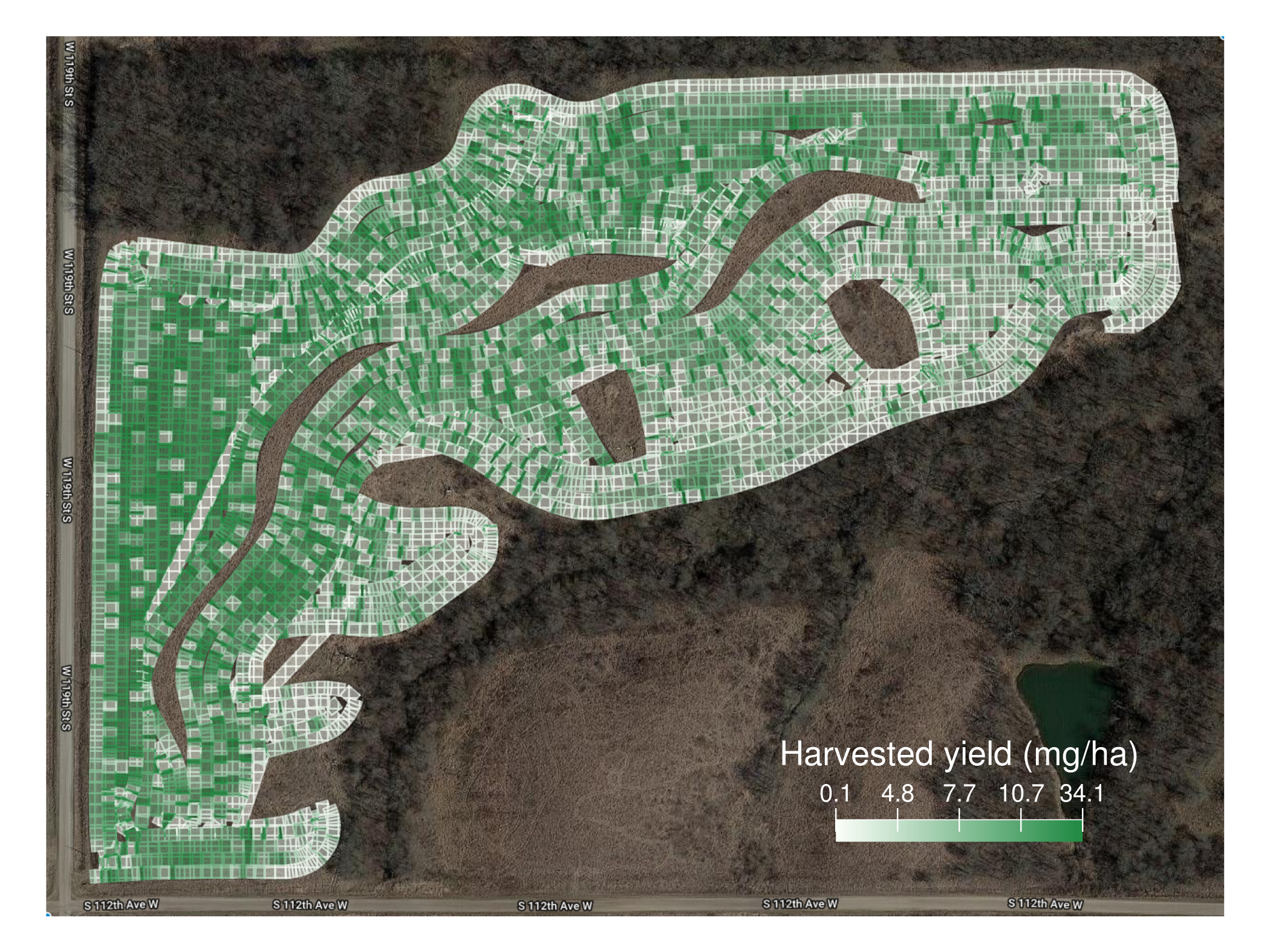}
    \caption{Tessellated plane partition}
   \end{subfigure}
  \begin{subfigure}[b]{0.49\textwidth}
    \centering
    \includegraphics[width=1\textwidth]{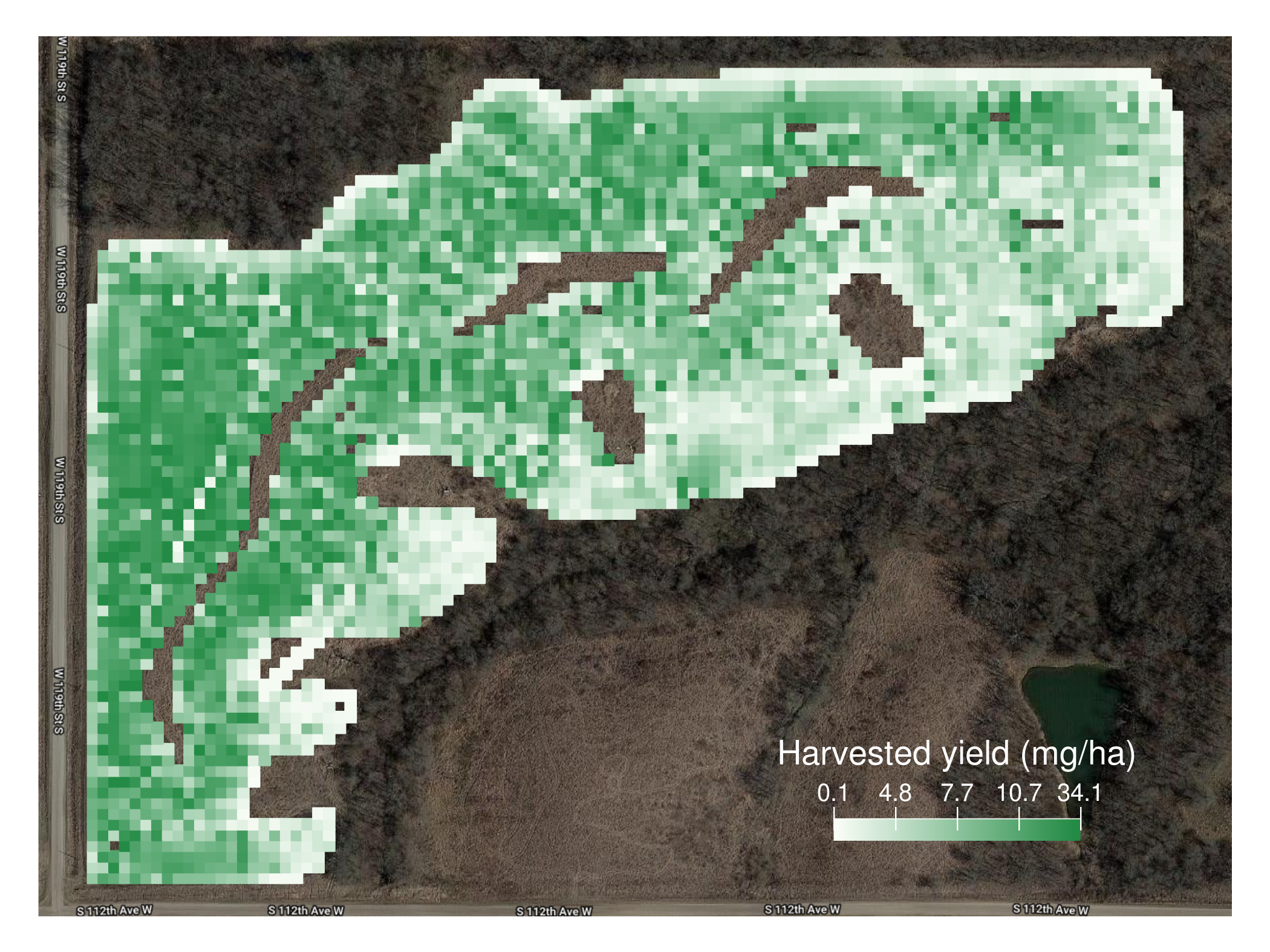}
    \caption{Apportioning step output}
  \end{subfigure}
  \begin{subfigure}[b]{0.49\textwidth}
    \centering
    \includegraphics[width=1\textwidth]{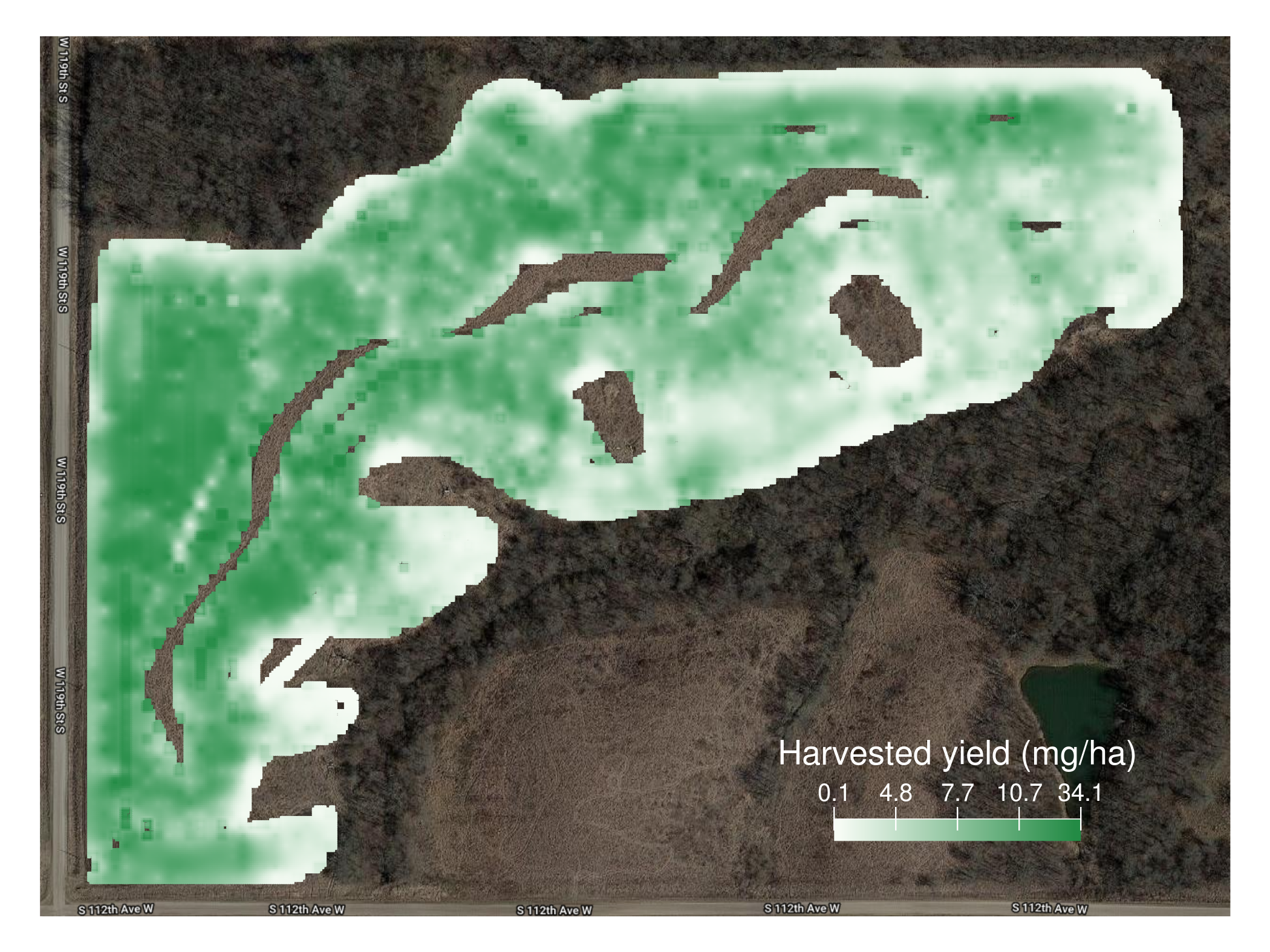}
    \caption{Smoothing step output}
  \end{subfigure}
  \caption[Step-by-step visualization of the algorithm for one
  field]{Step-by-step progression of Basswood (2012).}%
  \label{fig:basswood2012-all-steps}
\end{figure}

\end{document}